\documentclass[10pt,conference]{IEEEtran}
\IEEEoverridecommandlockouts
\usepackage{cite}
\usepackage{amsmath,amssymb,amsfonts}
\usepackage{algorithmic,graphicx,textcomp,xcolor}

\newcommand{\s}[1]{\, {\rm #1}} 
\usepackage{tikz}
\def\BibTeX{{\rm B\kern-.05em{\sc i\kern-.025em b}\kern-.08em
    T\kern-.1667em\lower.7ex\hbox{E}\kern-.125emX}}
\newcommand\copyrighttext{
    \small \begin{center} \color{gray} \textcopyright\,2026 IEEE.  Personal use of this material is permitted.  Permission from IEEE must be obtained for all other uses, in any current or future media, including reprinting/republishing this material for advertising or promotional purposes, creating new collective works, for resale or redistribution to servers or lists, or reuse of any copyrighted component of this work in other works. \end{center}}
\newcommand\copyrightnotice{
    \begin{tikzpicture}[remember picture,overlay]
    \node[anchor=south,yshift=10pt] at (current page.south) {\parbox{\dimexpr\textwidth-\fboxsep-\fboxrule\relax}{\copyrighttext}};
    \end{tikzpicture}
    }
\begin{document}

\title{Intrinsic neuro-synaptic spiking dynamics \\ and resonance in memristive networks}


\author{
    \IEEEauthorblockN{Yinhao Xu}
    \IEEEauthorblockA{\textit{School of Physics}\\
    \textit{University of Sydney}\\
    NSW, Australia\\
    yinhao.xu@sydney.edu.au}
\and
    \IEEEauthorblockN{Georg A. Gottwald}
    \IEEEauthorblockA{\textit{School of Mathematics and Statistics}\\
    \textit{University of Sydney}\\
    NSW, Australia\\
    georg.gottwald@sydney.edu.au}
\and
    \IEEEauthorblockN{Zdenka Kuncic}
    \IEEEauthorblockA{\textit{School of Physics}\\
    \textit{University of Sydney}\\
    NSW, Australia\\
    zdenka.kuncic@sydney.edu.au}
}
\maketitle
\copyrightnotice
\begin{abstract}
Self-organizing memristive networks are physical circuits that dynamically reconfigure their circuitry in response to external input signals.
Their adaptive behavior arises from intrinsic neuro-synaptic dynamics combined with a heterogeneous network topology. 
In this work, we demonstrate that such networks naturally generate neuronal population spiking dynamics similar to those observed in biological neuronal systems.
This study investigates the intrinsic and emergent dynamics of memristive networks mathematically and numerically for both DC and AC input signals. 
Nonlinear spike-like features are maximized when the frequency of the input driving signal matches the network's intrinsic dynamical timescale, where nonlinear resonance is observed. Furthermore, the optimal frequency for computation is found to be the maximal frequency before the onset of resonance.

\end{abstract}

\begin{IEEEkeywords}
Neuromorphic computing, Memristive dynamics, Nanowire networks
\end{IEEEkeywords}
\section{Introduction}
Neuromorphic computing aims to overcome the power limitations of conventional computing systems, particularly for artificial intelligence applications, by drawing direct inspiration from the brain's remarkable ability to efficiently perform complex information processing tasks \cite{Mehonic-Kenyon_2022,schuman2022opportunities, schmidgall_brain-inspired_2024, li2024brain, kudithipudi2025neuromorphic}.

One approach to neuromorphic computing aims to mimic the dynamics of biological neural networks (BNNs) on conventional computing platforms and complementary metal–oxide–semiconductor (CMOS) devices, leveraging off existing technologies and current advances in artificial neural networks (ANNs) to reach more efficient computation. 
Spiking neural networks (SNNs) are one such approach, where information is conveyed between artificial neurons via action potentials, or ``spikes'', a characteristic feature of BNNs \cite{maass1997networks, eshraghian2023training}. 
Since the biological brain is not yet well understood, alongside the high computational complexity involved in simulating realistic BNNs, complex network-dependent biological neuron dynamics are typically simplified to point neuron models, such as the commonly used leaky integrate-and-fire (LIF) model \cite{burkitt2006review}, which is sufficient to efficiently encode temporal structure and perform computation for SNNs while allowing for scalable simulation and CMOS hardware implementation.

Another promising approach to neuromorphic computing, is to leverage physical devices which naturally exhibit brain-like properties and intrinsic neuro-synaptic dynamics, without being constrained by biological fidelity. 
Memristive devices are one such example, as the memristors themselves naturally allow for synaptic plasticity and encoding of memory \cite{Song_2023}. 
When combined with the neuromorphic connectivity structure of self-organizing memristive networks, brain-like dynamics such as dynamical phase transitions (including avalanche criticality) and synchronization emerge \cite{mallinson2019avalanches,hochstetterAvalanches2021}. 
This approach is not constrained by either biology or current advances in ANNs and CMOS technology. Instead, information is processed naturally by the physics of the underlying substrate itself \cite{Jaeger_2023}, rather than by digital hardware, akin to analog computers and, unsurprisingly, the biological brain.

This study focuses on memristive nanowire networks, a self-organizing electrical circuit with axon-like nanowires and synapse-like memristive junctions at nanowire cross-points \cite{vahlBraininspired2024, kuncicNeuromorphic2020a, kuncicNeuromorphic2021}. 
Conductance changes in the memristive junctions are facilitated by electro-ionic transport, which is the source of all nonlinear network dynamics. 
These networks can thus be interpreted as high-dimensional nonlinear dynamical systems, as ``reservoirs'' to process information in the reservoir computing (RC) framework. 
Via RC, by linearly mapping the network voltage readouts to the desired output, memristive nanowire networks have been used for a wide range of information processing and learning tasks, including MNIST handwritten digit classification \cite{ zhu2023}, waveform regression \cite{hochstetterAvalanches2021, zhuInformation2021} and multivariate chaotic time series prediction \cite{xu2025, xuIJCNN2025}. 

This study demonstrates that the spiking behavior crucial to all neuron models and the functionality of SNNs, arises naturally in memristive nanowire networks, and shows how these nonlinear dynamical features can be optimized for physical RC tasks. 
Its origins are described mathematically and investigated numerically.
\section{Methods}

\subsection{Network Circuit Model}
The physical neuromorphic network simulated in this study is based on a model of self-assembled nanowire networks which exhibit neuro-synaptic nonlinear responses under electrical stimulation, governed by memristive junctions between nanowire cross-points \cite{kuncicNeuromorphic2021}.
The physical network can be abstracted as a graph, with nodes representing nanowires and edges representing memristive junctions, while the network connectivity is generated by simulating the bottom-up self-assembly process of the nanowires \cite{loeffler2020topological}.
The network must satisfy Kirchhoff's circuit laws at all times, which produces a system of linear equations that can be solved for the node voltages as a function of time; details of this nodal analysis technique can be found in previous studies \cite{zhuInformation2021, hochstetterAvalanches2021, baccettiErgodicity2024, xu2025}. 

Equivalently, the dynamic network circuitry can also be simulated via the edge voltages. 
Since all intrinsic dynamics originate from the memristive edges (see section~\ref{sec:mem_model}), this formulation allows for a simpler analysis of the network dynamics.
For any resistive network of $N$ nodes, with $E$ edges weighted by variable conductance $\mathbf{g}(t) \in \mathbb{R}^E$ at time $t$, let $G(t) = \text{\rm diag} (\mathbf{g}(t))$. 
It is then possible to define a projection operator $\Omega(G, t)$ that maps some input edge voltages $\mathbf{s}(t) \in \mathbb{R}^E$ to other edge voltages $\mathbf{v}(t)\in \mathbb{R}^E$ via $\mathbf{v}(t) = \Omega(G, t) \mathbf{s}(t)$ such that Kirchhoff's laws are satisfied. This projection operator is given by
\begin{equation}
	\Omega(G) = I - B(B^\top G B)^{-1}B^\top G \; ,\label{eq:Omega}
\end{equation}
where $B$ is the reduced incidence matrix (where columns corresponding to the input and ground nodes are removed to ensure linear independence of columns, while also serving as boundary conditions). 
Equation \eqref{eq:Omega} is a modified version of the $\Omega$ operator introduced by \cite{caravelli2017complex} such that it internally incorporates a conductance dependency.  

Note that experimentally, voltages can only be delivered into the circuit through the nodes. 
Let $\mathbf{\tilde{r}}(t) \in \mathbb{R}^N$ be the input voltages across the entire network, then the input edge voltages are given by $\mathbf{s}(t) = B^\top \mathbf{\tilde{r}}(t)$.

\subsection{Memristor Model} \label{sec:mem_model}
The memristor model is responsible for conductance evolution of $G(t)$ in the graph. 
The model considered here is a threshold model~\cite{kuncicNeuromorphic2020a,zhuInformation2021,hochstetterAvalanches2021}, where for a single memristor at edge $i$ with voltage $v_i(t)$, 
\begin{equation}
    \dot{x}_i  = \left\{
	\begin{array}{ll}
		(|v_i|-V_{\text{set}}) \;
        \text{sgn}(v_i)
        & \quad |v_i|>V_{\text{set}} \\
		0 & \quad V_{\text{reset}} \leq |v_i| \leq V_{\text{set}}\\
		q (|v_i| - V_{\text{reset}}) 
        \;\text{sgn} (x_i)
        & \quad V_{\text{reset}} > |v_i|\\
		0 & \quad |x_i|\geq x_{\rm max}
	\end{array}\right. . \label{eq:dx}
\end{equation}
Here, $\dot{x}_i(t)$ is the time derivative of $x_i(t)$, a state variable at the $i$th edge representing a flux linkage (in units of Volt seconds), with $q=10$ controlling the decay rate of memory, and $V_{\rm set}=0.01 \s{V}$, $V_{\rm reset}=0.005 \s{V}$, $ x_{\rm max} = 0.015 \s{Vs}$ are free parameters. 

\begin{figure}[h]
	\centering
	\includegraphics[width=0.95\linewidth]{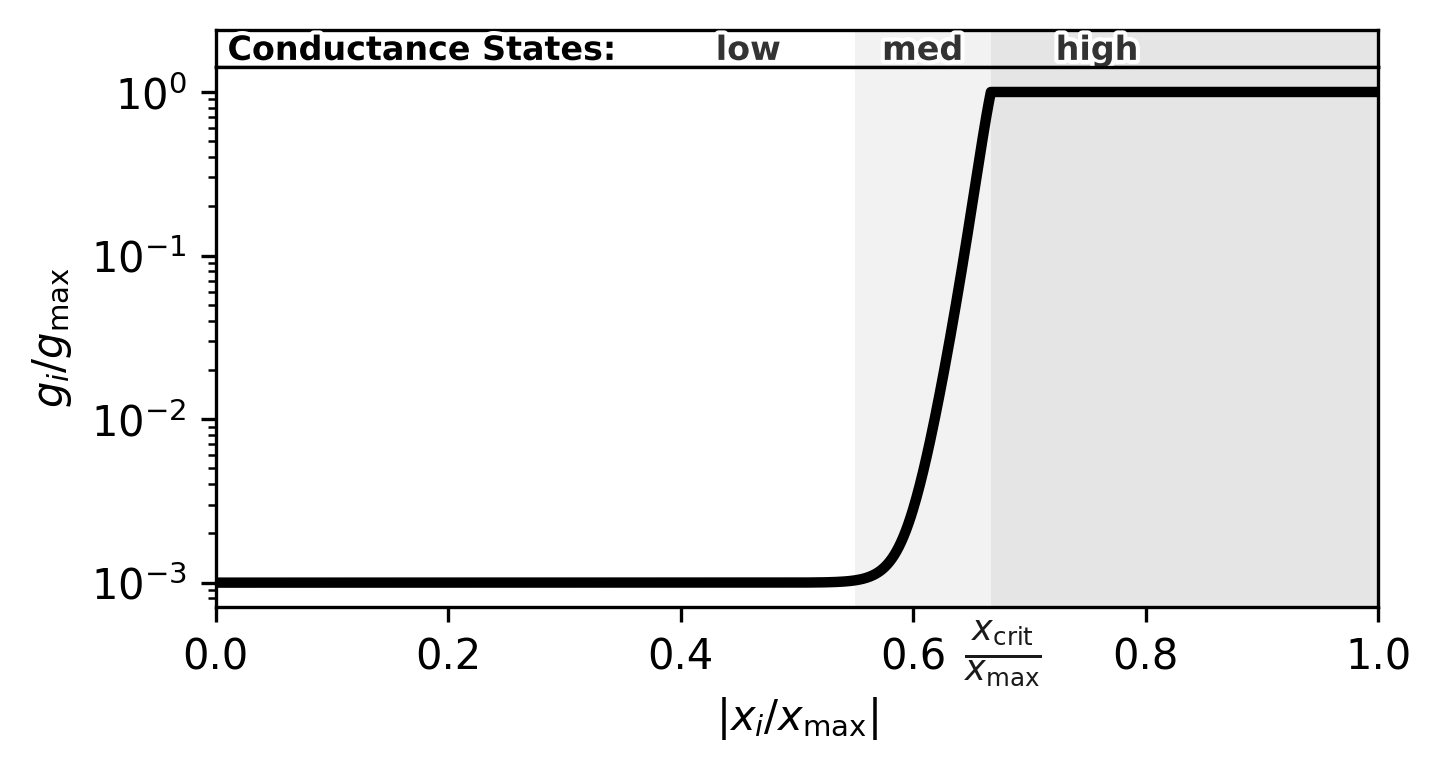}
	\caption{
    Normalized edge conductance as a function of the internal state parameter $x$, highlighting also 
    low, medium and high conductance states, with $x_{\rm crit}$ indicated on the boundary between medium and high conductance states. 
    }
	\label{fig:conductance}
\end{figure}

The conductance $g_i(x_i)$ at the $i$th edge is plotted in Fig.~\ref{fig:conductance} as a function of $x_i / x_{\rm max}$, which shows that $g_i(x_i)$ spans approximately three orders of magnitude, and rapidly increases once $x$ approaches $x_{\rm crit} = 0.01 \s{Vs}$, smoothened by a tunneling effect. More details of this memristor model can be found in \cite{hochstetterAvalanches2021, xu2025}.

\subsection{Network Dynamics}\label{sec:network_dynamics}
This section describes the collective dynamics that emerge at the network level.
Given edge voltages $\mathbf{v}(t)$ depend on conductance $\mathbf{g}(\mathbf{x}(t))$, with $\dot{\mathbf{x}}$ given by \eqref{eq:dx}, and $\mathbf{g}$ is known, the dynamics on ${\mathbf{v}} = \Omega(G, t) \mathbf{s}(t)$ is obtained by taking its time derivative and using \eqref{eq:Omega}, to give
\begin{equation}
    \dot{\mathbf{v}} = -B(B^\top G B)^{-1} B^\top \dot{G} \mathbf{v} + \Omega \dot{\mathbf{s}} \;.  \label{eq:vdot0}
\end{equation}
Introducing the $E\times E$ matrix $K(G) = B(B^\top G B)^{-1} B^\top$, the $E$-dimensional vector $\mathbf{u}(\dot{G},\mathbf{v}) =\dot{G}\mathbf{v}$ and the vector field $\mathbf{w}(G, \dot{\mathbf{s}}) = \Omega(G)\dot{\mathbf{s}}$, we can write \eqref{eq:vdot0} in the more compact form
\begin{equation}
    \dot{\mathbf{v}} =- K \mathbf{u} + \mathbf{w}.
    \label{eq:vdot}
\end{equation}
It is now clear that the voltage dynamics can be separated into two distinct components: an intrinsic memristive network dynamics from $K \mathbf{u}$ that depends on the network topology, and an extrinsic component externally driven by the input signal in $\mathbf{w}$. 

The vector field $\mathbf{w}$ drives all edge voltages to follow the dynamics of the input voltages $\mathbf{s}$, and does not necessarily generate nonlinear dynamics. 
It is trivial to see that if $G$ is time independent, then $K \mathbf{u}=0$ and $\dot{\mathbf{v}} = \mathbf{w} = \Omega \dot{\mathbf{s}}$, resulting in a linear network that is entirely driven by the external signal. 

Rich nonlinear dynamics are thus facilitated by the intrinsic network dynamics.
Consider the case of a DC input voltage signal, where $\dot{\mathbf{s}} = \mathbf{0}$; in this case, all observed dynamics originate from the network's intrinsic dynamics.
Here, the $i$th component of $\mathbf{u}(t)$ is $u_i(t) = \dot{g_i}(t) v_i(t)$, the voltage across edge $i$ weighted by its conductance time derivative. 
The matrix $K$ connects this vector to the network topology, where closer connected components are assigned a greater weighting, hence all changes in the edge voltages are a linear combination of $u_i(t)$ terms, weighted entirely by the network connectivity. 

Consequently, the network topology plays an important role in the network dynamics. 
On a dense network all $u_i(t)$ terms are weighted similarly by $K$ resulting in relatively little variety amongst edges, while on a sufficiently sparse network, $K$ acts similarly to some scalar multiple of the identity matrix, where no network effects would be apparent \cite{xuIJCNN2025}.

Examples of network dynamics are demonstrated and explained in the following section.
\section{Results \& Discussion}
\begin{figure}
	\centering
	\includegraphics[width=\linewidth]{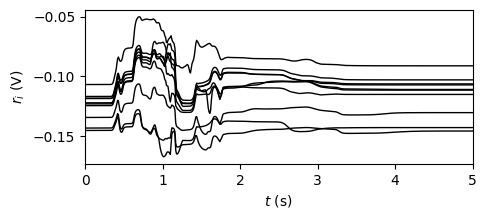}
	\caption{
    Ten arbitrary node readout voltages $r_i(t)$ from a simulated 100-node, 261-edge memristive network under a $0.2\s{V}$ DC bias applied over $5$ s.
    }
	\label{fig:node_spikes}
\end{figure}

Fig.~\ref{fig:node_spikes} shows the evolution of node voltage readouts $\mathbf{r}(t)$ from a 100-node 261-edge memristive network under a $0.2\s{V}$ DC input. 
This connectivity structure respects the network topology of the physical device \cite{loeffler2020topological}. 
The nodes exhibit action potential-like responses remarkably similar to those observed in cortical culture neuronal networks, measured across multiple channels \cite{Beggs-Plenz_2003}. 
The following section explains these collective network dynamics via the theory presented in section~\ref{sec:network_dynamics}.
The implications for information processing and neuromorphic computing will also be explicated below in the context of physical RC. 
\subsection{Spike-like Dynamics}\label{sec:spike}
\begin{figure*}
	\centering
	\includegraphics[width=0.9\linewidth]{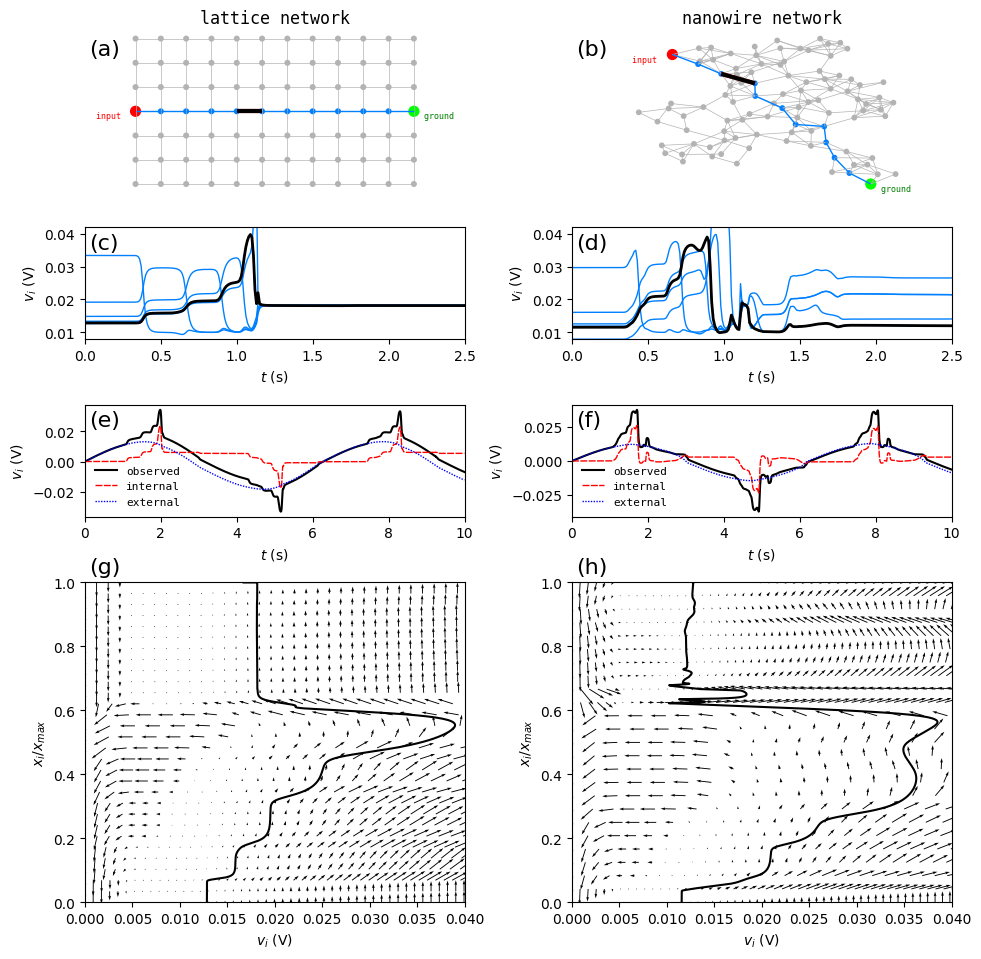}
	\caption{Memristive network dynamics simulated on (a) a uniform lattice network and (b) a heterogeneous nanowire network (based on the connectivity of the physical substrate). An input voltage signal is delivered via the input (red) node and current flows toward the ground (green) node. 
    The edges which form a high conductance path are highlighted in blue, and a reference edge (black) is used to investigate local dynamics in the sub-figures.
    Spike-like responses are evident in (c-d) for the two networks under a $0.2 \s{V}$ DC bias, and also in (e-f), which shows responses under a $0.16 \s{Hz}$, $0.2 \s{V}$ AC signal. 
    The memristive network dynamics are plotted in phase space in (g-h), with reference trajectories (black) matching the voltage curves in (c-d).
    }
	\label{fig:spikes}
\end{figure*}

From \eqref{eq:vdot}, a constant DC input signal corresponds to $\dot{\mathbf{s}}=0$ and hence, the action potential-like ``spikes'' observed in Fig.~\ref{fig:node_spikes} are induced solely via the network's internal self evolution. 
To better understand the nature of these spike-like dynamics, Fig.~\ref{fig:spikes} demonstrates the effect of the network connectivity.

Fig.~\ref{fig:spikes}(a) and Fig.~\ref{fig:spikes}(b) show graphical representations of two memristive network topologies: a uniform lattice and a heterogeneous network, based on a nanowire network model. Each memristive network is set up as a 2-terminal device, with one node representing a designated input (red) and one node representing the ground (green). Blue edges highlight the conductance path that subsequently forms under electrical stimulation and the black edge highlights a reference edge selected for the analyses shown in subplots (c)--(h).

On a lattice network, action-potential-like responses to a DC input signal are evident in the edge voltages shown in Fig.~\ref{fig:spikes}(c). 
In terms of \eqref{eq:vdot}, the voltage changes here are a linear combination of other edge voltages (weighted by $K$) which may have already completed their voltage self-evolution and spiked at earlier times.
Relative to the reference edge (black curve), a few nearby edge voltages on the conductance path are also shown in blue in Fig.~\ref{fig:spikes}(c), which verifies the above claim, as their response times match the step-like changes observed in the reference edge. 
The voltage differences drop rapidly after all neighboring edges have completed their conductance evolution, after which $\dot{g_i}=0$ and hence $\dot{v_i}=0$. 
The uniformity of the lattice network produces a relatively clean spiking profile which aids with interpretation, but for a more realistic heterogeneous network (cf. Fig.~\ref{fig:spikes}(b)), complex network interactions lead to more varying and diverse voltage dynamics, as seen in Fig.~\ref{fig:spikes}(d). 

Fig.~\ref{fig:spikes}(e-f) show corresponding edge voltage profiles when an AC input signal is applied to each network, revealing remarkably similar network dynamics. 
From \eqref{eq:vdot}, the dynamics can be split into internal and external components, facilitated by $K \mathbf{u}$ and $\mathbf{w}$ respectively.
It can be seen that the internal network dynamics is the sole contributor to the spike-like response, and behaves similar to the DC case shown in Fig.~\ref{fig:spikes}(c-d). 
The observed voltage response (black) is the sum of the voltages constructed from the external input-driven dynamics (blue) and the internal memristive network dynamics (red).

The observed expressive dynamics arises from a complex interplay between the network topology, $\mathbf{x}$ and past voltage states, all encoded in \eqref{eq:vdot0}. 
From \eqref{eq:dx} it is apparent that $\dot{x}_i \propto v_i$, hence a larger voltage input increases the change in conductance and thus voltage evolution via \eqref{eq:vdot0}, though as seen below, they still roughly follow similarly shaped trajectories in the phase space of $x$ and $v$. 
Fig.~\ref{fig:spikes}(g-h) shows the internal network dynamics $K \mathbf{u}$ for the reference edge as a vector field in phase space between the state parameter $x_i(t)$ and edge voltage $v_i(t)$ for the lattice and nanowire networks respectively, with the reference trajectory (black) that matches Fig.~\ref{fig:spikes}(c-d). 
Here, a complete picture of the memristive dynamics can be seen, where for sufficiently high voltage (larger than $V_{\rm set}=0.01\s{V}$) for any $x$, the voltage traverses through phase space such that a spike-like response naturally forms, followed by a rapid decrease before relaxing to a lower constant voltage, finally reaching $x_{\rm max}$.  
From this, the memristive network spiking dynamics can be explained as an emergent phenomenon of conductance path formation, where by Fig.~\ref{fig:spikes}(g-h) for the spike to repeatedly occur this conductance path must be broken again (i.e. $|x_i|\ll x_{\rm crit}$) before supplying this edge with sufficient voltage signal. 

The action potential-like responses demonstrated here differ from those of biologically inspired point neuron models, most notably the LIF model and its variants, which generally do not model nonlinear network-dependent conductance changes \cite{burkitt2006review}.
In the neuromorphic memristive network presented here, action potential-like responses arise naturally and self-consistently from the intrinsic dynamics and conductance path formation with any sufficiently large voltage input; no manual interference or mathematical models are required. 
Furthermore, all dynamics on the neuromorphic network originate from memristive dynamics on the edges which propagate throughout the network, as described by \eqref{eq:vdot}. 
Consequentially, every spike-like response in the memristive network is also made up of previous neighboring spiking events, as seen in Fig.~\ref{fig:spikes}(c). 
In contrast, point neuron models impose nonlinear dynamics entirely within the neuron itself, and although the network can affect the neurons via synaptic input signals, they do not change its internal nonlinear dynamical state. 

Interestingly, however, memristive elements are implicated in the more biologically realistic Hodgkin-Huxley axon circuit model \cite{hodgkin1952quantitative, Chua_HH-memristor_2012}, where action potentials are driven by changes in the axon membrane potential, due to the opening and closing of ion channels. 
This bears strong similarity to the memristive nanowire networks considered here, where under an electrical bias, ions migrate across membrane-like junctions between axon-like nanowires. 
The key difference here is that the resulting action potential-like responses are shaped by network interactions; just like neurons in a neuronal network, nanowires in a nanowire network do not operate in isolation, but rather in cooperation with their connected neighbors. This corroborates independent insights from computational neuroscience that indicate spike dynamics are generated by network dynamics rather than by a Poisson process \cite{Gerstner2002spiking}. 


\subsection{Resonance and Computation}
The results presented thus far suggest that the spike-like nonlinear responses in memristive networks may be considered an example of emergent dynamics, which is supported by previous studies \cite{Diaz2019emergent, hochstetterAvalanches2021}. 
In the context of computation, and specifically physical RC, this suggests computational performance depends on the expressivity and prevalence of these nonlinear features. 
As shown in section.~\ref{sec:spike}, maximizing the occurrence of the nonlinear features is effectively equivalent to maximizing the occurrence of spike-like responses.
Fig.~\ref{fig:spikes}(c-f) and \eqref{eq:vdot0} both demonstrate that for a short time after the spiking event, the internal network-driven dynamics rapidly decays regardless of whether the input signal is DC or AC, ending such nonlinear responses. 
Under a continuous AC signal (cf. Fig.~\ref{fig:spikes}(e-f)), it may be possible to identify an optimal driving frequency such that the spike-like features occur as frequently as possible while minimizing down-time, and not be too frequent such that it destructively interferes with the intrinsic dynamical timescale of $x(t)$ evolution (i.e. not enough time for conductance path formation).

\begin{figure}
	\centering
	\includegraphics[width=0.9\linewidth]{figs/resonance.png}
	\caption{Average amplitude of spike-like responses (averaged across $1000 \s{s}$) as a function of frequency and amplitude of an input AC signal.
    Simulated on a 100-node 261-edge memristive nanowire network.
    }
	\label{fig:resonance}
\end{figure}

Evidently, if the external oscillation is applied to match the network's intrinsic frequency response, then the resulting voltage spikes will peak at higher amplitudes than if the external input signal is applied at other frequencies. 
This can be used to identify the resonant frequencies in Fig.~\ref{fig:resonance}, which shows the average voltage amplitude (across $1000 \s{s}$) of spike-like responses as a function of frequency and amplitude of the input AC signal. The frequencies corresponding to the maximal spike amplitudes are the resonant frequencies for that particular input amplitude. 
Note that since there is no closed-form analytic solution to \eqref{eq:vdot}, the
dynamical timescale and hence resonant frequency of this network can only be determined empirically. 
Here, the resonant frequency ranges between $0.2$ to $1 \s{Hz}$ depending on the amplitude of the input AC signal.
The voltage spikes also peak at higher amplitudes with increasing input amplitude, and resonance effects can only be clearly observed at input amplitudes larger than $0.3 \s{V}$. 

While these results bear strong resemblance to resonance effects in a resonant circuit or a driven damped simple harmonic oscillator, the memristive network model does not assume an RLC circuit or other oscillators. 
Rather, the origin of this resonance effect may be attributed to the collective memory of the memristive edges and their nonlinear dynamics, which introduce an intrinsic dynamical timescale associated with excitation and relaxation, facilitated by the closed-loop feedback loop between $\mathbf{v}$, $\mathbf{g}$, and $\mathbf{x}$ as described by \eqref{eq:vdot} -- when excitation dominates over relaxation, resonance may occur. 

\begin{figure}
	\centering
	\includegraphics[width=0.9\linewidth]{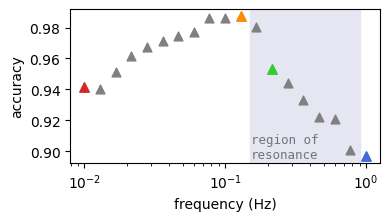}
	\caption{Physical RC performance in a waveform regression task, where a $0.5\s{V}$ sine wave is input to a 2-terminal 100-node, 261-edge memristive network reservoir, and readout voltage features are fit to a triangle wave of the same frequency.
    Accuracy is 1-RNMSE (where RNMSE is root-normalized-mean-squared error).
    The shaded area corresponds to the region of resonance observed in Fig.~\ref{fig:resonance} for $0.5\s{V}$ input amplitude. 
    The colored markers correspond to the trajectories in Fig.~\ref{fig:resonance_phase}.
    }
	\label{fig:nonlin_trans}
\end{figure}

Having established how to maximize the occurrence of the nonlinear features generated by the network (in the form of spikes), computational performance as a function of those features is considered next.
Fig.~\ref{fig:nonlin_trans} shows the accuracy of a simple RC task where an input sine wave ($0.5\s{V}$ amplitude) is delivered to one node of a nanowire network reservoir and voltage features of remaining nodes are used to linearly regress to a triangle wave of the same frequency, which requires addition of higher harmonics generated by the nonlinear spikes.
In Fig.~\ref{fig:resonance} for $0.5 \s{V}$ input amplitude (black), the largest frequency prior to resonance is around $0.13\s{Hz}$, which coincides with the optimal driving frequency observed for this waveform regression task. 

\begin{figure}
	\centering
	\includegraphics[width=0.8\linewidth]{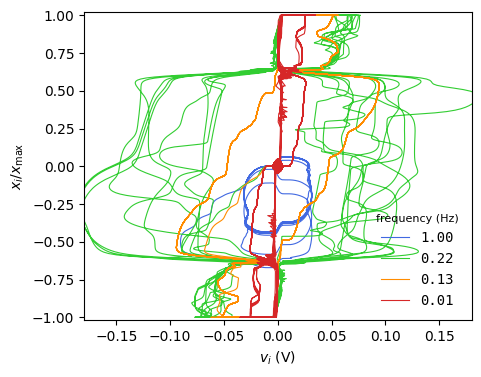}
	\caption{The trajectories of $x_i$ and $v_i$ through phase space for an example reference edge (same as that of Fig.~\ref{fig:spikes}(b)), with $0.5\s{V}$ sine wave inputs of various frequencies to a 2-terminal 100-node, 261-edge memristive network. 
    The colors match the colored markers in Fig.~\ref{fig:nonlin_trans}.
    }
	\label{fig:resonance_phase}
\end{figure}

Interestingly, accuracy drops rapidly at higher frequencies, which may be attributed to the onset of chaotic dynamics. 
Fig.~\ref{fig:resonance_phase} depicts trajectories in $x$ and $v$ phase space for an example memristive edge utilized for this task, which shows chaotic trajectories within the region of resonance for $0.22 \s{Hz}$ (green). 
At higher frequencies beyond resonance, such as for $1\s{Hz}$ (blue) in Fig.~\ref{fig:resonance_phase}, $\mathbf{x}(t)$ no longer has sufficient time to fully traverse the phase space, 
 hence nonlinear effects are stymied. 
As shown in a previous study \cite{hochstetterAvalanches2021}, memristive nanowire networks can be driven into a chaotic dynamic regime by a high frequency and high amplitude AC signal and such a regime is detrimental to learning. 
As such, the optimal frequency for computation can now be interpreted as the frequency that allows for the maximal occurrence of nonlinear dynamics (in the form of spike-like responses) before the network becomes chaotic. 
\section{Conclusion}
This study demonstrates that action potential-like nonlinear responses arise naturally in memristive networks from their intrinsic neuro-synaptic network-dependent dynamics.
These spike-like features can be considered as an emergent manifestation of Kirchhoff constraints on the nonlinear network dynamics. Moreover, the theoretical description outlined in this study shows a clear delineation between intrinsic network dynamics and extrinsic dynamics due to the input signal, providing an explanation for the qualitatively similar dynamical spike-like features under different input driving signals.

The results also reveal that the occurrence and amplitude of these spike-like dynamics are maximized when the frequency of the input signal matches the network's intrinsic dynamical timescale associated with its memory elements and connectivity.
Implementation of a physical reservoir computing task showed that performance is optimized close to this resonance-like dynamical regime, which corresponds to a frequency that drives the maximal occurrence of nonlinear spike-like dynamical features before the network exhibits resonance.

\section*{Acknowledgment}
Y.X. is supported by an Australian Government Research Training Program (RTP) Scholarship.
G.A.G. acknowledges support from the Australian Research Council under Grant No. DP220100931.

\bibliographystyle{ieeetr} 
\bibliography{references}

\end{document}